\documentclass[default,iicol]{sn-jnl}

\usepackage{natbib}

\jyear{2021}%

\theoremstyle{thmstyleone}%
\theoremstyle{thmstyletwo}%

\theoremstyle{thmstylethree}%

\begin{document}

\title[Article Title]{Extreme-ultraviolet structured beams via high harmonic generation}


\author*[1]{\fnm{Alok Kumar} \sur{Pandey}}\email{alok-kumar.pandey@etu.univ-amu.fr}

\author*[2]{\fnm{Alba de las} \sur{Heras}}\email{albadelasheras@usal.es}

\author[2]{\fnm{Julio} \sur{San Román}}

\author[2]{\fnm{Javier} \sur{Serrano}}

\author[2]{\fnm{Luis} \sur{Plaja}}

\author[1]{\fnm{Elsa} \sur{Baynard}}

\author[1]{\fnm{Moana} \sur{Pittman}}

\author[3]{\fnm{Guillaume} \sur{Dovillaire}}

\author[1]{\fnm{Sophie} \sur{Kazamias}}

\author[4]{\fnm{Charles} \sur{G. Durfee}}

\author[2]{\fnm{Carlos} \sur{Hernández-García}}

\author[1]{\fnm{Olivier} \sur{Guilbaud}}

\affil*[1]{\orgdiv{Laboratoire Irène Joliot-Curie, Université Paris-Saclay, UMR CNRS, Rue Ampère, Bâtiment 200, F-91898, Orsay Cedex, France}} 

\affil*[2]{\orgdiv{Grupo de Investigación en Aplicaciones del Láser y Fotónica, Departamento de Física Aplicada, Universidad de Salamanca, E-37008 Salamanca, Spain}} 

\affil[3]{\orgdiv{Imagine Optic, 18, rue Charles de Gaulle, 91400 Orsay, France}} 

\affil[4]{\orgdiv{Department of Physics, Colorado School of Mines, Golden, Colorado 80401, USA}}





\abstract{Vigorous efforts to harness the topological properties of light have enabled a multitude of novel applications. Translating the applications of structured light to higher spatial and temporal resolutions mandates their controlled generation, manipulation,  and thorough characterization in the short-wavelength regime. Here, we resort to high-order harmonic generation (HHG) in a noble gas to upconvert near-infrared (IR) vector, vortex, and vector-vortex driving beams that are tailored respectively in their Spin Angular Momentum (SAM), Orbital Angular Momentum (OAM), and simultaneously in their SAM and OAM. We show that HHG enables the controlled generation of extreme-ultraviolet (EUV) vector beams exhibiting various spatially-dependent polarization distributions, or EUV vortex beams with a highly twisted phase. Moreover, we demonstrate the generation of EUV
vector-vortex beams (VVB) bearing combined characteristics of vector and vortex beams.
We rely on EUV wavefront sensing to unambiguously affirm the topological charge scaling of the HHG beams with the harmonic order. Interestingly, our work shows that HHG allows for a synchronous controlled manipulation of SAM and OAM. These EUV structured beams bring in the promising scenario of their applications at nanometric spatial and sub-femtosecond temporal resolutions using a table-top harmonic source.}




\maketitle 

\section{Introduction}\label{intro}

Recent advances in harnessing the topological properties of light beams have led to exciting new developments in various fields \cite{Vector-vortex-Ref1-Forbes, Dunlop-review}. Out of numerous complex and exotic configurations \cite{Vector-vortex-Ref1-Forbes}, vortex and vector beams are perhaps among the paramount examples of structured light. On the one hand, vortex beams carrying Orbital Angular Momentum (OAM) manifest azimuthally twisting wavefront given by $exp (i\ell \phi)$ around the beam propagation axis \cite{OAM-Allen}, where the topological charge $\ell$, also known as the phase winding number, signifies the number of $2\pi$ shifts across the azimuthal coordinate $\phi$ in the transverse plane. On the other hand, vector beams are characterized by their spatially varying polarization, which 
is encoded in their spin angular momentum (SAM) \cite{vector-vortex-Ref4-Vector-Zhan}. A vector beam can also be interpreted as a superposition of two optical vortices with counterrotating circular polarization (right circular RCP and left circular LCP) and equal but opposite topological charges, thus null OAM: $\ell_{RCP} = -\ell_{LCP}$. Consequently, the phase difference between the two circularly polarized components governs the spatially varying linear polarization distribution. Contrastingly, vector-vortex beams (VVB) that are tailored simultaneously in their SAM and OAM, can be interpreted as a superposition of two circularly polarized optical vortices of opposite handedness and distinct topological charges: $\ell_{RCP} = \ell_{p} + 1$ and $\ell_{LCP} = \ell_{p} - 1$, where the topological Pancharatnam charge $\ell_{p}$ defined in terms of the Pancharatnam-Berry phase \cite{vector-vortex-Ref46} accounts for the intertwined phase and polarization properties of the VVB \cite{vector-vortex-Ref38}. 

Structured light beams have enabled numerous interdisciplinary applications \cite{Vector-vortex-Ref1-Forbes, Dunlop-review, vector-vortex-Ref4-Vector-Zhan}. In particular, structured light has opened new frontiers in light-matter interactions: photonic induction and control of ultrafast currents in semiconductors, thus opening the possibilities of reconfigurable optoelectronic circuits \cite{semiconductor-current1, semiconductor-current2}, OAM transfer to valence electrons \cite{OAM-bound-electron}, photoelectrons \cite{OAM-free-electron}, observation of magnetic helicoidal dichroism \cite{Ruchon-3}, and selective excitation of multipolar modes in localized surface plasmons \cite{Multipolar-surface-plasmon}, to name a few. Given the plenitude of applications, controlled generation and manipulation of light beams exhibiting OAM, SAM, or both in the short-wavelength regime are particularly relevant as they can allow extending their applications to nanometric spatial and subfemtosecond temporal scales \cite{Garcia-Review2}. We note that short-wavelength structured beams have been demonstrated at large-scale synchrotron and X-ray free-electron lasers facilities utilizing diffractive optical elements \cite{OAM-FEL, OAM-FEL2}. However, these alternatives often suffer from poor throughput,  lack of widespread accessibility, and less tunability in the sense that each desired configuration requires a custom-designed diffractive optical element. On the other hand, high-order harmonic generation (HHG) resulting from the highly-nonlinear interaction of intense laser light with atoms \cite{vector-vortex-Ref48-Schafer-ATI}, leads to highly coherent, ultrashort, and high-frequency (odd multiples of the driving field) radiation rainbow that can extend up to soft-X-ray spectral range \cite{Popmintchev1}. On the macroscopic level, when the emitted radiation from a large number of individual atoms is phase-matched, the HHG allows mapping certain characteristics of the driving beam into the harmonic radiation. Exploiting the field-driven coherent nature of the HHG process, long-wavelength structured-beam driven HHG in noble gases has made long strides, enabling control and manipulation of OAM and/or SAM in the extreme-ultraviolet (EUV) spectral range \cite{Garcia2013, Gariepy, Geneaux, David-article, vector-vortex-Ref66, vector-vortex-Ref68, Self-Torque, vector-vortex-Ref53, Garcia-Vector, Garcia-Fractional, Optica-vector-vortex, ACS-Photonics-Pandey2022, Sanson1}.

In this work, we resort to HHG in a 15 mm long argon-filled gas cell to demonstrate the generation and characterization of EUV structured beams. We show that the HHG process allows for a continuous tunability of the polarization state of the upconverted vector beam from radial to azimuthal. Moreover, by driving HHG with IR vortex beams, we report the production, and single-shot intensity, wavefront characterization of EUV vortex beams exhibiting very high topological charges.  Furthermore, we demonstrate that the HHG process allows combining the spatially inhomogeneous polarization distribution of a vector beam and the twisted wavefront of vortex beams in a controlled manner, yielding EUV vector-vortex beams carrying large OAM, and tunable polarization state. Controlled generation and manipulation of EUV light with phase and/or polarization singularities pave the way for their applications in fundamental and applied studies using a table-top high-harmonic source.

\section{Experimental setup}

\begin{figure*}[ht]
\centering
\includegraphics[width=0.9\linewidth]{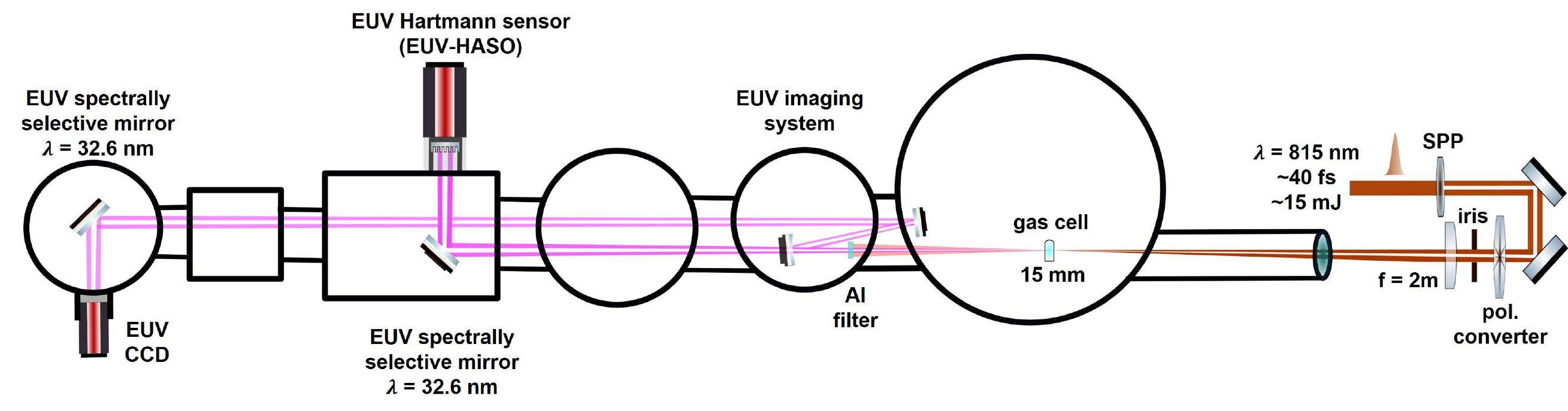}
\caption{Experimental setup for the generation and characterization of structured HHG beams. To obtain the IR vector driver, a segmented polarization converter (pol. converter) is inserted in the incoming vertically-polarized fundamental Gaussian beam. The polarization converter and a spiral phase plate (SPP) are concatenated to yield the vector-vortex driving beam. The IR driving beam is focused into a 15 mm long argon-filled gas cell by a 2 m focal length lens to generate high harmonics. The remaining IR driver is removed using a 300 nm thick Al filter. The harmonic beam is guided towards either the EUV Hartmann wavefront sensor (EUV-HASO) located $\sim1.98\: \rm m$ from the HHG source or the nearfield imaging system. Spectral and polarization filtering by the $45^{0}$ multilayer plane mirror allows detection of the vertical polarization component of the $25^{\rm th}$ harmonic beam centered at $\lambda_{EUV} = $ 32.6 nm in the far field of HHG source (at EUV HASO) and the exit of generation medium.}
\label{vector-vortex-setup}
\end{figure*}

In Figure \ref{vector-vortex-setup}, we depict the experimental setup to generate EUV vector and VVB from a linearly polarized near-infrared (IR) Gaussian beam of central wavelength 815 nm, pulse duration $\sim$40 fs, $\sim$15 mJ maximum energy, and diameter $\sim$24 mm at $1/e^{2}$. On the one hand, the vertically polarized incoming Gaussian beam is transformed into a vector beam using a large aperture (3-inch diameter) segmented polarization converter composed of eight half-wave plates with azimuthally varying optical axis orientation. On the other hand, the IR vector-vortex driver with topological charge $\ell_{p1}$ is obtained by inserting both a spiral phase plate (SPP) and the polarization converter in the incoming Gaussian beam. We remark that the utilized configuration allows continuous tuning of the polarization state of the vector and vector-vortex driving beams from radial to azimuthal by altering the polarization of the fundamental IR beam from vertical (s-polarization) to horizontal (p-polarization) using a half-waveplate.


We focus the IR vector, vortex, or vector-vortex driving beam into a 15 mm long argon-filled gas cell using a 2-meter focal length lens to generate high harmonics. The remaining IR driving beam after the generation medium is filtered using a 300 nm thick Al filter. Thereafter, we steer the HHG structured beam towards a high-resolution EUV Hartmann wavefront sensor (EUV-HASO, Imagine Optic) through a narrowband 45-degree multilayer flat mirror \cite{XUV-Mirror}. This narrowband flat mirror serves for two purposes: spectral and polarization filtering of the structured HHG beams. The vertical polarization intensity component of the $q = 25^{\rm th}$ harmonic of the driving beam centered at $\lambda_{EUV} = $ 32.6 nm is guided to the EUV wavefront sensor: the extinction of neighboring orders exceeds 90\% \cite{XUV-Mirror}, while the experimentally determined polarization selectivity is better than $\sim$10:1. As we will show later, our approach based on spectral and polarization filtering allows for an unambiguous interpretation of the topological charge and/or polarization state of the EUV structured beams. Additionally, a high-magnification ($\sim$6) spectrally selective imaging system facilitates the characterization of the vertical polarization intensity component of HHG beams at the exit plane of the generation medium. The EUV imaging system is composed of a narrowband near-normal incidence ($\sim1.2^{0}$) multilayer converging mirror of focal length $\sim500$ mm. Though this configuration is not sensitive to the polarization due to the small angle of incidence, we image the exit plane of the gas cell through the polarizing 45-degree multilayer flat mirror placed before the EUV CCD (see Fig. \ref{vector-vortex-setup}). This allows us to characterize the vertical polarization intensity component of the harmonic beam at the gas cell exit plane, hence permitting an unambiguous affirmation of the HHG beams’ polarization state at the source. As the used EUV multilayer mirrors offer high reflectivity within a small spectral range, we use a spectral-phase control of the IR driver by an acousto-optic modulator (Dazzler, FASTLITE) to spectrally tune the high-harmonic beam. We remark that the truncation of a vector beam can deform its structure upon propagation \cite{Vector-truncation1, Vector-truncation2}. Therefore, in the results presented here, we avoid the aperturing of IR vector and vector-vortex driving beams (see Fig. \ref{vector-vortex-setup}). On the other hand, in the case of HHG driven by IR vortex, we aperture the incoming beam by an iris of 18 mm diameter to optimize the phase matching.  

\section{Results and discussion}\label{results}

\subsection{EUV vector beams via HHG}

Owing to their remarkable properties, particularly under high numerical aperture focusing conditions \cite{vector-vortex-Ref16, Garcia-vector-2}, vector beams have enabled a wide range of novel applications \cite{vector-vortex-Ref4-Vector-Zhan, vector-vortex-Ref16, Gupta, Payeur}. Conclusively, obtaining intense vector beams in the EUV domain is highly desirable. We also note that the frequency upconversion of vector beams in a thin (500 $\rm \mu m$ gas jet) generation medium was reported in \cite{Garcia-Vector}, where authors used a Rowland-circle type spectrometer as a polarizer to affirm the vectorial polarization of HHG beams. Here, we use a 15 mm long gas cell and exploit our EUV imaging system (Fig. \ref{vector-vortex-setup}) to unambiguously confirm the spatially variant polarization distribution of the HHG beam.

\begin{figure*}[ht]
\centering
\includegraphics[width=0.95\linewidth]{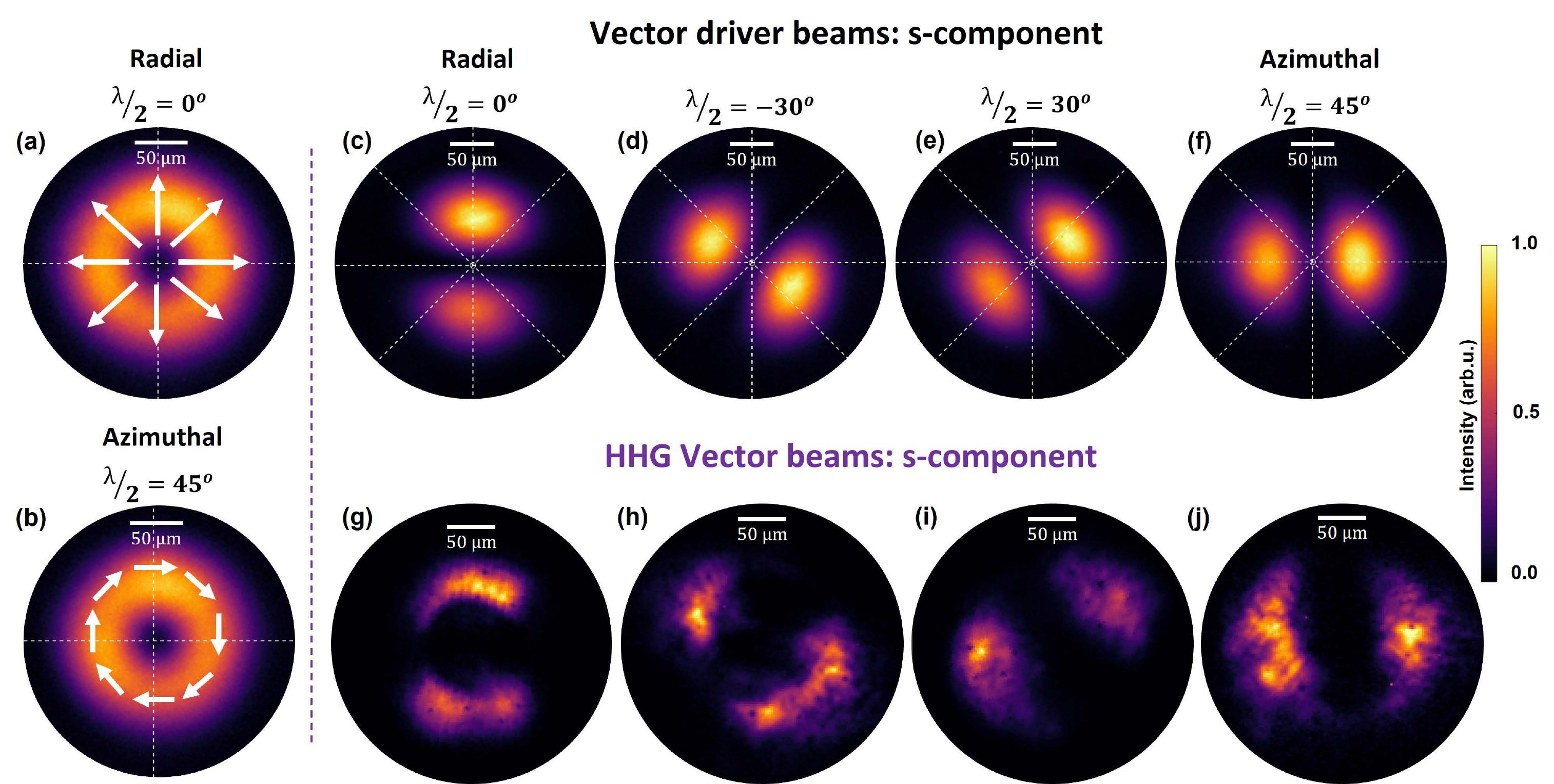}
\caption{Polarization-structured EUV vector beams via HHG. Left panel: experimental intensity distribution of the IR vector beams with (a) radial and (b) azimuthal polarization distributions at the waist. Right panel: top rows shows the vertical polarization (‘s’-polarization) component of the driving beam bearing radial (c), azimuthal (f), and intermediate polarization states (d, e). Note that the white arrows in the left panel represent tentative polarization distribution, which is affirmed through the intensity profiles of vertically polarized components depicted in the right panel. In the bottom row, we show the vertical polarization components of the HHG vector beams ($25^{\rm th}$ harmonic) manifesting radial (g), azimuthal (j), and in-between polarization distributions (h, i). These intensity distributions of HHG vector beams are acquired using the monochromatic polarization selective EUV imaging system that is configured to image the exit plane of the gas cell (see Fig. \ref{vector-vortex-setup}).}
\label{HHG-vector-beam}
\end{figure*}

In Figure \ref{HHG-vector-beam}, we show the measured intensity distribution of radially (a) and azimuthally (b) polarized IR vector beams at the waist. In both cases, the intensity distribution presents an annular structure with a dark hole at the center. We note that in contrast to the vortex beams exhibiting a null on-axis intensity due to helical wavefront \cite{OAM-Allen}, the donut-like profile of a vector beam arises from on-axis polarization singularity resulting from spatially varying linear polarization \cite{vector-vortex-Ref4-Vector-Zhan}. We mark the tentative polarization distribution using white arrows in Fig. \ref{HHG-vector-beam}(a, b). To affirm their polarization state, we image the vertical polarization intensity component of the IR vector beams. Indeed, the two vertical intensity lobes for radial (Fig. \ref{HHG-vector-beam}(c)), and the two horizontal lobes for azimuthal (Fig. \ref{HHG-vector-beam}(f)), confirm their respective polarization states. As noted before, by rotating the polarization of the fundamental IR Gaussian beam from vertical to horizontal using a $\lambda/2$ wave plate, we can continuously vary the polarization distribution from radial ($\lambda/2 = 0^{0}$) to azimuthal ($\lambda/2 = 45^{0}$). In Figure  \ref{HHG-vector-beam}(d, e), we present the vertical polarization intensity distributions for two intermediate polarization states: $\lambda/2$ wave plate is rotated by (d) $\sim-30^{0}$, and (e) $\sim30^{0}$ from the neutral axis. In all the cases, the two-lobed intensity profile signifies their vectorial polarization distribution, while their orientation represents radial, azimuthal, or intermediate polarization states.  


To generate high-harmonics, the IR vector driving beam is focused into the Argon gas cell, as depicted in Fig. \ref{vector-vortex-setup}. Subsequently, the vertical polarization intensity component of the $25^{\rm th}$ harmonic at the gas cell exit plane is acquired by employing the polarization-selective monochromatic imaging system (see Fig. \ref{vector-vortex-setup}). In the bottom row of Fig. \ref {HHG-vector-beam}, we show the intensity distribution of the vertical polarization component of HHG vector beams exhibiting radial (g), azimuthal (j), and intermediate polarization distributions (h, i). Remarkably, akin to the vertical polarization components of IR vector beams (Fig. \ref{HHG-vector-beam}(c-f)), the HHG beams also display a two-lobed profile whose orientation closely follows the trend of the IR driving beams. Conclusively, the coherent nature of the HHG process indeed permits the generation of EUV beams with vectorial polarization distribution in an extended generation medium (15 mm gas cell) from the IR vector drivers. A further loose focusing geometry and a longer gas cell can therefore allow the production of intense EUV vector beams, bringing in the prospect of their practical applications in the EUV domain. Additionally, we note that HHG allows upscaling of the photon energies up to the soft X-ray wavelength range \cite{Popmintchev1}, hence further adding to the utility of HHG-based polarization structured short-wavelength pulses. We also remark that unlike frequency upconversion in nonlinear crystals where spatially varying linear polarization of the fundamental vector beam is unfavorable, the uniqueness of the HHG process permits the generation of EUV vector beams.

\subsection{Generation of high-harmonic vector-vortex beams}

\begin{figure*}[ht]
\centering
\includegraphics[width=0.95\linewidth]{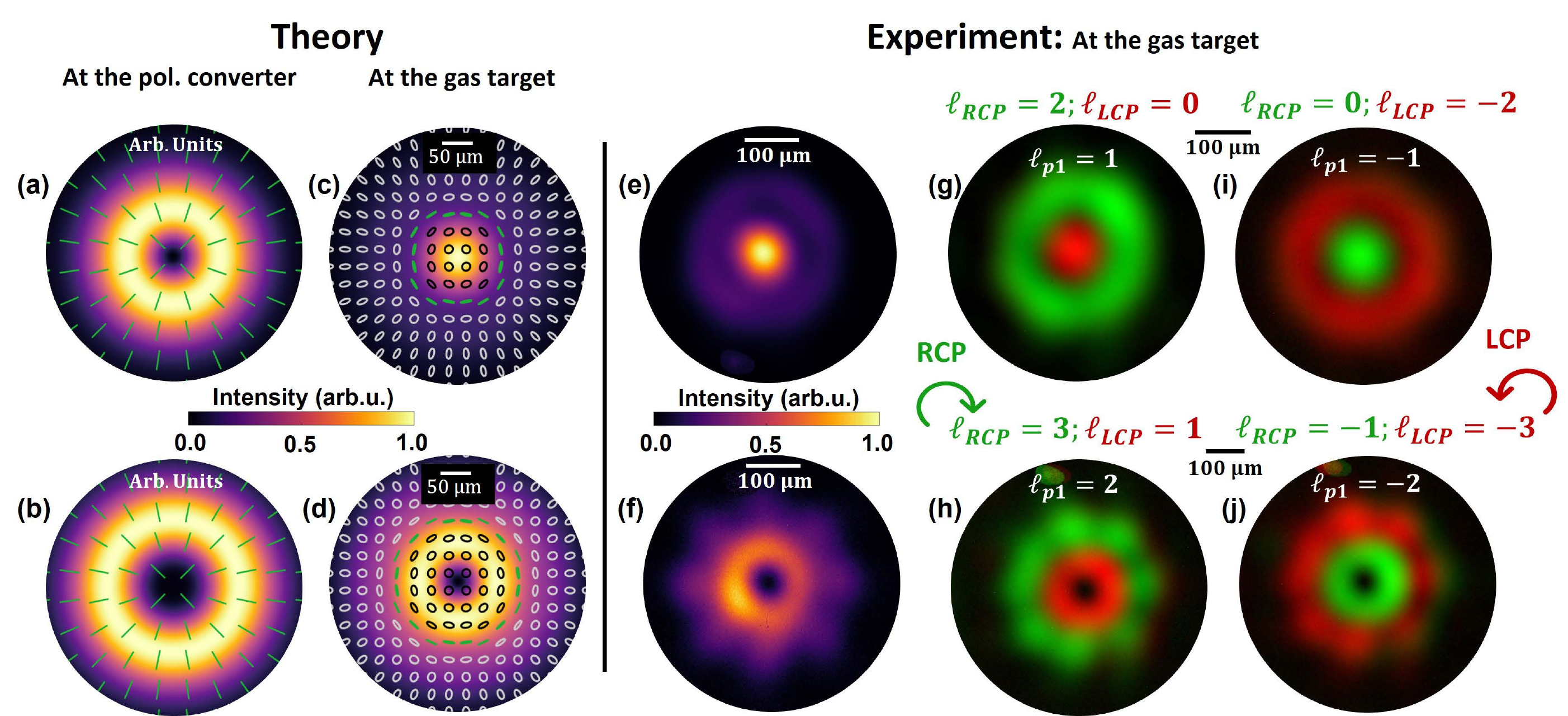}
\caption{Theoretical (left panel) and experimental (right panel) characterization of the IR vector-vortex driver of topological charges $\lvert \ell \rvert = 1$ (top row) and $\lvert \ell \rvert = 2$ (bottom row). For radially polarized driving beams, we show the theoretical intensity maps with embedded polarization ellipses right after the polarization converter (a, b) and after propagating to the gas cell (c, d). Note that the clean input beam (a, b) with linear radial polarization transforms into a complex polarization distribution at the gas target (c, d). The right panel depicts the experimental profiles (e, f), and the separated RCP and LCP components at the gas target for IR VVB of topological charge (g) $\ell_{p1} = +1$,  (i) $\ell_{p1} = -1$, (h) $\ell_{p1} = +2$, and (j) $\ell_{p1} = -2$. The helicity of the circular polarization between the two parts that compose the total beam reverses when the sign of topological charge is changed from positive (g, h) to negative (i, j).}
\label{IR-vector-vortex}
\end{figure*}

Here, we detail the high-harmonic frequency upconversion of IR VVB driving beams of different topological Pancharatnam charges that are obtained by concatenating SPP and the polarization converter (see Fig. \ref{vector-vortex-setup}). In Figure \ref{IR-vector-vortex}, we depict the theoretical (left panel) and experimental (right panel) characterization of the IR vector-vortex driving beam of $\lvert \ell_{p1}\rvert = 1$ (top row) and $\lvert \ell_{p1}\rvert = 2$ (bottom row). The theoretical panel shows the evolution of the intensity profile and the polarization distribution—denoted by the embedded polarization ellipses—as the driving beam propagates from the polarization converter (a, b) to the gas cell (c, d).  In the experimental panel, in addition to the intensity distribution of the IR driver at the waist (e, f), we present the separated RCP and LCP components comprising the total beam for the respective cases in (g, h). The topological charge of the RCP and LCP components is inferred by using the relationship between the topological charge and the radius of maximum intensity \cite{R-max}. We note that the divergence of Laguerre-Gauss modes depends on the topological charge \cite{Divergence-Padgett}. As the two components comprising the total beam exhibit different topological charges, vector-vortex beams manifest complex behavior upon propagation. In particular, the radial polarization right after the polarization converter (a, b) transforms into a complex polarization distribution resembling a full-Poincaré beam \cite{vector-vortex-Ref40, vector-vortex-Ref44}. Nevertheless, a ring of linear polarization (indicated by the green lines in (c, d)) occurs wherever RCP and LCP components overlap with equal intensity. Interestingly, in the region of linear polarization, the dephasing induced between the RCP and LCP components due to the difference in their Gouy phase leads to the evolution of the initial polarization distribution from radial at the polarization converter (a, b) to azimuthal at the gas target (c, d). Contrastingly, in the case of a pure vector beam comprising two counterrotating circularly polarized vortices with equal and opposite topological charges ($\ell_{RCP} = -\ell_{LCP}$), there is no polarization rotation due to the absence of phase shift.  This aspect is evident from the experimental characterization presented in Fig. \ref{HHG-vector-beam}(c to f), where the initial polarization state remains static as the IR vector beams propagate towards the waist after the focusing lens.


\begin{figure*}[ht]
\centering
\includegraphics[width=1\linewidth]{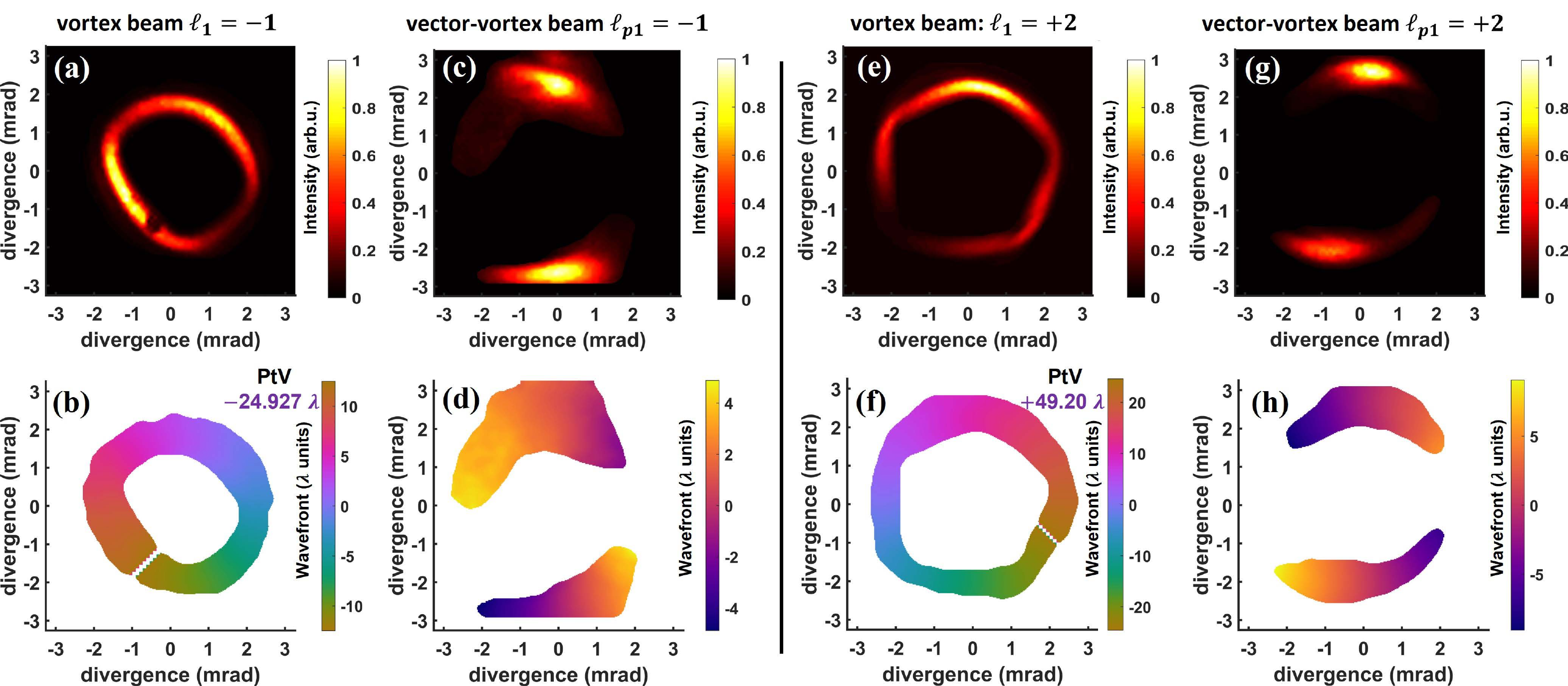}
\caption{Characterization of high-OAM HHG vortex and vector-vortex beams. Intensity (top row) and wavefront (bottom row) profiles of the $25^{\rm th}$ harmonic beam for IR vortex driver of (a, b) $\ell_{1} = -1$ and (e, f) $\ell_{1} = +2$.  In (c, d) and (g, h), we depict the vertical polarization intensity and wavefront components of the HHG beam for radially polarized vector-vortex drivers of $\ell_{p1} = -1$ and $\ell_{p1} = +2$, respectively. In comparison to harmonic vortex beams with uniform linear-vertical polarization, the two-lobed profiles with vertical orientation in (c, d) and (g, h) represent spatially inhomogeneous linear polarization with the radial distribution. Additionally, azimuthally twisting wavefront in both the lobes with clockwise (h) and anticlockwise (d) rotation for the positive and negative sign of the topological charge indicates the control of the OAM helicity of harmonic VVB.}
\label{vector-vortex-l2-rad-pol}
\end{figure*}

To drive HHG, the IR VVB is focused into the gas cell (see Fig. \ref{vector-vortex-setup}). In Figure \ref{vector-vortex-l2-rad-pol}, we depict the far field vertical polarization intensity (top row) and wavefront (bottom row) profiles of the HHG VVB driven by the IR beam of topological Pancharatnam charge $\ell_{p1} = -1$ (c, d), and $\ell_{p1} = +2$ (g, h). To contrast the behavior concerning harmonic vortex beam exhibiting uniform vertical polarization, we present the intensity and wavefront maps of the HHG-OAM beam for IR driver of topological charge $\ell_{1} = -1$ (a, b) and $\ell_{1} = +2$ (e, f). Note that all the wavefronts here are represented in the unit of the central wavelength of EUV light ($\lambda_{EUV} = 32.6 \: \rm nm$). Regarding uniformly polarized HHG vortex beams, we observe an annular intensity distribution (a, e) and azimuthally twisting wavefront (b, f). Interestingly, the $25^{\rm th}$ harmonic beam bears a total peak-to-valley (PtV) wavefront variation of $\sim -24.9 \lambda_{EUV}$ (b) and $\sim +49.2 \lambda_{EUV}$ (f) for the IR vortex driver of $\ell_{1} = -1$ and $\ell_{1} = +2$, respectively. Therefore, the total wavefront twist in the transverse plane indicating the topological charge of vortex beams is within $2\%$ of the theoretical expectation. These results unambiguously affirm the topological charge scaling of the HHG-OAM beams with harmonic order \cite{Garcia2013, ACS-Photonics-Pandey2022}: $\ell_{q} = q\ell_{1}$. Moreover, the sense of wavefront rotation changes from anticlockwise (b) to clockwise (f) following the sign of the topological charge of the driving beam, indicating tunability of OAM helicity for HHG-OAM beams. We also remark that the intensity and wavefront of the harmonic vortex beams are reconstructed from single-shot acquisitions, therefore showing that an extended generation medium allows the production of intense HHG-OAM beams without degradation of their high-charge vortex structure.

In contrast to the uniformly polarized EUV vortex, the vertical polarization intensity projection of the HHG vector-vortex beams displays a two-lobed profile with vertical orientation, indicating spatially inhomogeneous polarization distribution in the farfield of the HHG source. Note that though the driving beam at the gas cell plane exhibits complex polarization distribution (Fig. \ref{IR-vector-vortex}(c, d)), the severe restriction of harmonic efficiency on the ellipticity of the driving beam restricts the efficient generation in the region with low ellipticities \cite{vector-vortex-Ref72}. We remark that as the initial radial polarization at the polarization converter transforms to azimuthal at the focus of the driver (see Fig. \ref{IR-vector-vortex}(a-d)), the high-harmonics are generated initially with azimuthal polarization at the gas cell. However, as the Gouy phase-induced phase shift is the same for the IR driving and HHG beams \cite{Optica-vector-vortex},  the azimuthal polarization at the gas target remarkably transforms to near-radially polarized high-order harmonics in the far-field. Consequently, in the farfield of the HHG source (Fig. \ref{vector-vortex-l2-rad-pol}(c, d) and Fig. \ref{vector-vortex-l2-rad-pol}(g, h)), the vertical polarization projections exhibit two vertically oriented lobes. Furthermore, in both cases, the wavefront manifests an azimuthal twist whose direction of rotation follows the sign of the topological Pancharatnam charge of the IR driver. On the one hand, these results demonstrate that the harmonic beam indeed manifests combined characteristics of vector and vortex beams. On the other hand, they also reveal that the HHG process allows controlling the OAM helicity of the harmonic VVB.  

\begin{figure*}[ht]
\centering
\includegraphics[width=0.9\linewidth]{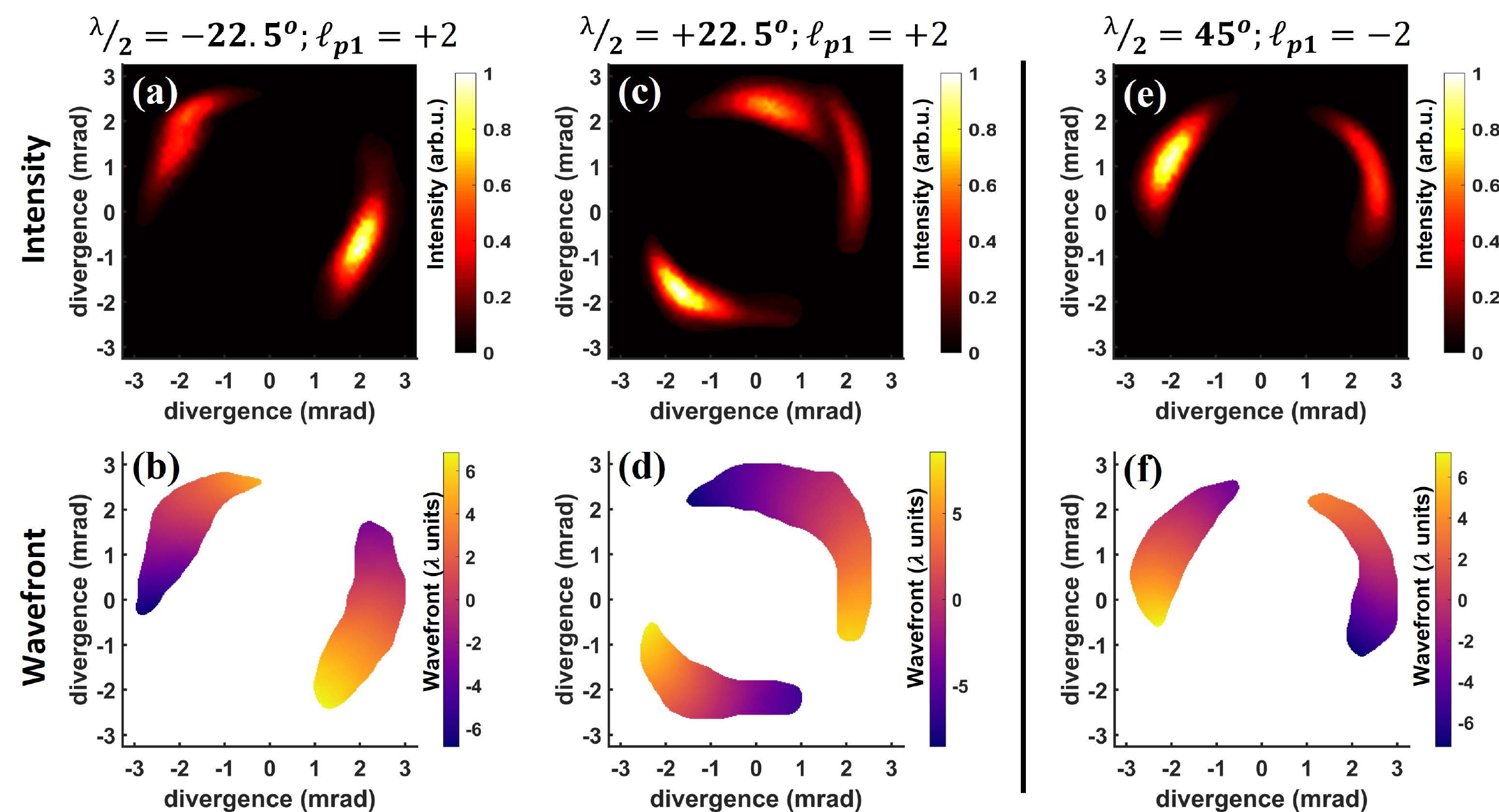}
\caption{HHG vector-vortex beams driven by $\ell_{p1} = +2$ and $\ell_{p1} = -2$ IR beam with intermediate and azimuthal polarization states. Left panel: measured intensity (top row) and wavefront (bottom row)  of the HHG vector-vortex beams with polarization states in between radial and azimuthal for $\ell_{p1} = +2$ IR driver. Note that the IR driving beams with intermediate polarization states are generated by varying the polarization of the fundamental Gaussian beam using a half-waveplate.  The orientation of the half-waveplate is noted on top of each column: $0^{0}$ corresponds to the neutral axis that leaves the initial vertical polarization of the Gaussian beam unchanged. The vertical polarization intensity distributions of the HHG beam show a rotated two-lobed profile, while the wavefronts manifest a clockwise twisting wavefront. Right panel: (e) intensity, and (f) wavefront of vertical component for azimuthally polarized HHG VVB driven by $\ell_{p1} = -2$ IR beam. Remarkably, the handedness of the wavefront twist reverses from clockwise (b, d) to anticlockwise (f) when the sign of topological charge changes from positive (b, d) to negative (f).}
\label{vector-vortex-l2-inter-pol}
\end{figure*}

We remark that as for HHG vector beams (Fig. \ref{HHG-vector-beam}), we were able to generate EUV VVB with intermediate polarization states by varying the polarization of the fundamental IR beam, as depicted in the left panel of Fig. \ref{vector-vortex-l2-rad-pol}. For azimuthally polarized IR driver of $\ell_{p1} = -2$, the right panel shows the vertical polarization components of (e) intensity and (f) wavefront. In all the cases, vector and vortex characteristics of the HHG beams are apparent from the two-lobed intensity distribution and the azimuthally twisting wavefront, respectively. Besides, as expected, the direction of wavefront rotation reverses with a change in the sign of $\ell_{p1}$ (Fig. \ref{vector-vortex-l2-rad-pol}(f)). In conclusion, the experimental results in Figs. \ref{vector-vortex-l2-rad-pol}, \ref{vector-vortex-l2-inter-pol} demonstrate the controlled generation of HHG VVB, and manipulation of their polarization state, as well as, their OAM helicity.

\subsubsection{Topological charge scaling of HHG VVB}

In the case of complete intensity distribution, as for uniformly-polarized harmonic vortex beams (Fig. \ref{vector-vortex-l2-rad-pol}(a, b) and Fig. \ref{vector-vortex-l2-rad-pol}(e, f)), the topological charge can be conveniently retrieved from the overall wavefront twist along the azimuthal coordinate. However, as the HHG vector-vortex beams exhibit a non-uniform linear polarization distribution,  the spectrally selective EUV mirror at $45^{0}$ angle of incidence offer maximum reflectivity for the vertically polarized component, while the horizontal polarization component is poorly reflected (a factor of 10:1). Consequently,  the HHG vector-vortex beams detected by the wavefront sensor present a two-lobed intensity distribution (Figs. \ref{vector-vortex-l2-rad-pol}, \ref{vector-vortex-l2-inter-pol}). Therefore, to deduce the topological charge in these cases, we adopt a different approach. We subtract the theoretically defined phase of different topological charges from the experimental wavefront map. Thereafter, the theoretical order that yields a minimum root-mean-square (RMS) wavefront residue designates the topological charge of the HHG VVB.

\begin{figure*}[ht]
\centering
\includegraphics[width=0.8\linewidth]{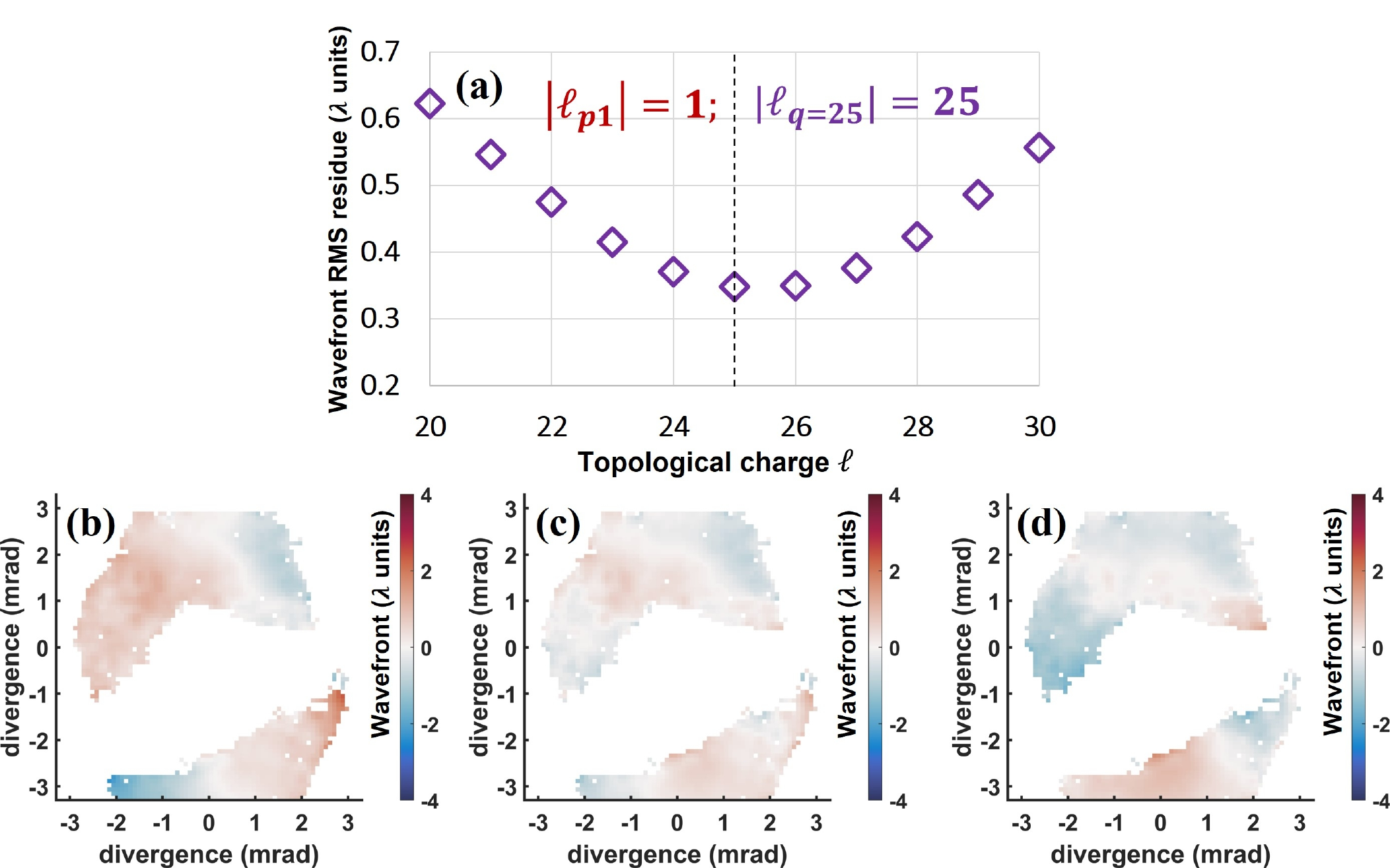}
\caption{Deducing the topological charge of experimental HHG vector-vortex beams. (a) For radially polarized HHG vector-vortex driven by $\ell = -1$ IR beam, we show the RMS wavefront residue in $\lambda$ units after subtracting the theoretical phase of various topological charges. A minimum RMS wavefront residue is obtained for $\ell = 25$, which coincides with the theoretically formulated scaling of Pancharatnam topological with the harmonic order: $\ell_{pq} = q\ell_{p1}$. Residual phase after subtracting (b) $\ell = 20$, (c)  $\ell = 25$, and (d) $\ell = 30$.}
\label{vector-vortex-charge}
\end{figure*}

We present the result of this analysis for the selected case of HHG vector-vortex driven by the radially polarized beam of $\ell_{p1}  = -1$ in Fig. \ref{vector-vortex-charge}.  Remarkably, a minimum RMS wavefront residue is obtained after subtracting the theoretical azimuthal phase of order $\ell  = 25$ (Fig. \ref{vector-vortex-charge}(a)). To further corroborate this aspect, we show the residual wavefronts after subtracting the theoretical orders of $\ell = 20$, $\ell = 25$, and $\ell = 30$ in Figs. \ref{vector-vortex-charge}(b-d), respectively. Indeed, relatively, the flattest wavefront with a minimum RMS residue is obtained after subtracting the theoretical phase of topological charge $\ell = 25$ (comparing Figs. \ref{vector-vortex-charge}(b-d)). This behavior shows that for vector-vortex driven HHG, the topological charge of the $q^{\rm th}$ harmonic scales linearly with the Pancharatnam topological charge of the IR driving beam \cite{Optica-vector-vortex}: $\ell_{pq} = q\ell_{p1}$. Indeed, by considering separate conservation of SAM and OAM while satisfying parity conservation that restricts the number of IR photons participating in the HHG process to an odd integer, we can show that HHG VVB results from the superposition of two circularly polarized vortices with opposite helicity and distinct topological charges \cite{vector-vortex-Ref77, Optica-vector-vortex, Garcia-Fractional}: $\lvert \ell_{q, \: RCP} - \ell_{q, \: LCP}\rvert = 2$. For the IR driver of Pancharatnam topological charge $\ell_{p1} = -1$, the harmonic VVB is composed of: $\ell_{q, \: RCP} = q\ell_{p1} - 1$ and $\ell_{q, \: LCP} = q\ell_{p1} + 1$. Therefore, the HHG Pancharatnam charge, which is the average of the topological charge of RCP and LCP components, is equal to $-25$ in this case.  On the one hand, these results demonstrate that the selection rule for EUV vectorial-vortices obtained through HHG is governed by the upscaling of the topological Pancharatnam charge and not the OAM of the constituent RCP and LCP modes \cite{Optica-vector-vortex}. On the other hand, they simultaneously reflect on the merit of our experimental approach based on EUV wavefront sensing for the characterization of HHG structured beams.

\section{Conclusion}\label{CONCLUSIONS}

We demonstrate frequency upconversion of near-infrared structured beams to the EUV spectral range via HHG in a 15 mm long generation medium. Our work shows that HHG facilitates the generation of EUV vector beams exhibiting spatially variant linear polarization with tunable distribution ranging from radial to azimuthal. Concerning OAM-driven HHG, we demonstrate production and complete spatial characterization of intensity and wavefront of EUV vortex beams carrying topological charge up to 50. Owing to the long generation medium and thus a larger emitting volume, the signal level was sufficiently high to permit single-shot intensity and wavefront characterization of high topological charge HHG vortex beams. The characterization method based on EUV wavefront sensing allowed us to unambiguously assess the topological charge and OAM helicity of the HHG vortex beams, affirming the linear scaling of HHG beam topological charge with harmonic order \cite{Garcia2013, ACS-Photonics-Pandey2022}. Finally, through synchronous control of SAM and OAM of the driving beam, we show that the HHG process allows controlled generation of vector-vortex beams, thus merging the spatially varying polarization of a vector beam and the twisted wavefront of vortex beams. Moreover, the presented results reveal the possibility of tuning not only the topological charge but also the polarization state and the OAM helicity of the EUV VVB. Through wavefront characterization of HHG VVB, we experimentally demonstrate that the selection rule for EUV vectorial-vortices obtained through HHG is governed by the upscaling of topological Pancharatnam charge and not the OAM of the constituent RCP and LCP modes.

Ultrafast structured EUV light further widens the scope of table-top high-harmonic sources for fundamental and applied studies. On the one hand, tightly focused high topological charge ultrafast EUV vortex beams carrying large OAM may enable light-matter OAM transfer to atoms and molecules \cite{vector-vortex-Ref34}, and photonic induction of skyrmionic defect in magnetic materials \cite{Skyrmion-OAM}. On the other hand, intense EUV vector beams obtained through HHG in an extended generation medium pave the way for short-wavelength lithography \cite{Garcia-vector-67, Garcia-vector-68}, microscopy \cite{Garcia-vector-65}, and diffraction imaging applications \cite{Garcia-vector-25,Garcia-vector-66}. Our further work involves investigating the tight focusing of these structured EUV beams using a Kirkpatrick-Baez focusing system. Concerning HHG VVB with combined characteristics of vector and vortex beams, our work opens the possibility of synthesizing attosecond light-springs \cite{ Garcia2013} tailored to exhibit spatially varying polarization \cite{Optica-vector-vortex}, therefore offering a new degree of freedom in attosecond beams.

\backmatter

\bmhead{Acknowledgments}
The project leading to this publication has received funding from the ERC under the European Union’s Horizon 2020 research and innovation program (ATTOSTRUCTURA - grant agreement No 851201). We acknowledge the computer resources at MareNostrum and the technical support provided by Barcelona Supercomputing Center (FI-2020-3-0013). The authors thank CEMOX installation at IOGS, Palaiseau, France, for the design and fabrication of the multilayer optics. We acknowledge the technical support of IJC Lab staff J. Demailly and O. Neveu. We are thankful to F. Quéré (Institut rayonnements matière du CEA, Saclay) for providing the polarization converter used in this study. \newline

\noindent \textbf{Funding:} European Research Council (851201); Ministerio de Ciencia de Innovación y Universidades, Agencia Estatal de Investigación and European Social Fund (PID2019-106910GB-I00, RYC-2017-22745); Junta de Castilla y León and FEDER Funds (SA287P18); Université Paris-Saclay (2012-0333TOASIS, 50110000724-OPTX, PhOM REC-2019-074-MAOHAm); Conseil Régional, Île-de-France (501100003990); Barcelona Supercomputing Center (FI-2020-3-0013). \newline

\noindent \textbf{Data availability:} The raw data reported in this study are available from the corresponding authors upon request. \newline

\noindent \textbf{Competing interests:} The authors declare no competing interests.

\bibliographystyle{naturemag}
\bibliography{sn-bibliography}


\end{document}